\documentstyle[12pt,epsf,psfig]{article}
\begin{document} 
\title{\bf  Constraints of mixing matrix elements in the  sequential fourth 
generation model}  
\author{Wu-Jun Huo\\ {\sl
Institute of High Energy Physics, Academia Sinica, P.O. Box $918(4)$},\\{\sl
 Beijing $100039$, P.R.  China}}

\date{} 
\maketitle

\begin{abstract}

 We review our works on the sequential fourth generation model and
focus on the constriants of $4\times 4$ quark  mixing matrix
elements.  We investigate the quark mixing matrix elements from the rare $K,B$
meson decays. We talk about the $ hierarchy$ of the
$4\times 4$ matrix and the existence of fourth generation.   
\end{abstract}
\newpage

\section{Introduction}

 The Standard Model (SM) is a very successful theory of the elementary
particles known today. But it must be incomplete because it has too many 
unpredicted parameters ($ninteen!$) to be put by hand. Most of these 
parameters are in the fermion part of the theory. We don't know the source 
of the quarks and leptons, as well as
how to determinate their mass and number theoretically. We have to get their 
information all from experiment. There is still no successful 
theory which can be
 descripted them with a unified point, even if the Grand Unified 
 Theory\cite{grand} and Supersymmetry\cite{supers}.
Perhaps {\it elementary particles} have substructure 
 and we need to progress more elementary theories. But this is beyond our 
 current experimental level. 
 On the other hand, the recent
measurment of the muon anomalous magnetic moment by the experiment E821
\cite{e821} disagrees with the SM expectations at more than 2.6$\sigma$ level.
 There are  convincing evdences that
neutrinos are massive and oscillate in flavor \cite{neutrino}. 
It seems to indicate the presence of new physics.

 From the point of phenomenology,
for fermions, there is a realistic question is number of the fermions generation or
weather there are other additional quarks or leptons. The present experiments 
can tell us there are only three generation fermions with $light$ neutrinos
which mass are less smaller than $M_Z /2$\cite{Mark} but the experiments don't 
exclude the existence of other additional generation, such as the fourth
generation, with a $heavy$ neutrino, i.e. $m_{\nu_4} \geq M_Z /2$\cite{Berez}.
Many refs. have studied models which extend the fermions part, such as
vector-like quark models\cite{vec-like}, sterile neutrino models\cite{sterile}
and the sequential four generation standard model (SM4)\cite{McKay} which 
we talk in this note. We consider
 a sequential fourth generation non -SUSY model\cite{McKay}, which is added
  an up-like quark $t^{'}$, a down-like quark $b^{'}$, a lepton $\tau^{'}$, 
  and a heavy neutrino $\nu^{'}$ in the SM. The properties of these new 
  fermions are all the same as their corresponding counterparts 
  of other three generations except their masses and CKM mixing, see tab.1,
  
\begin{table}[htb] 
\begin{center} 
\begin{tabular}{|c||c|c|c|c|c|c|c|c|} 
\hline 
& up-like quark & down-like quark & charged lepton &neutral lepton \\ 
\hline 
\hline 
& $u$ & $d$& $e$ & $\nu_{e}$ \\ 
SM fermions& $c$&$s$&$\mu$&$\nu_{\mu}$ \\ 
& $t$&$b$&$\tau$&$\nu_{\tau}$\\
\hline
\hline new
fermions& $t^{'}$&$b^{'}$&$\tau^{'}$&$\nu_{\tau^{'}}$ \\ 
\hline 
\end{tabular}
\end{center}
\caption{The elementary particle spectrum of SM4} 
\end{table}

In SM4, the qurak mixing matrix can be wretten as,
\begin{equation}
V = \left (
\begin{array}{lcrr}
V_{ud} & V_{us} & V_{ub} & V_{ub'}\\
V_{cd} & V_{cs} & V_{cb} & V_{cb'}\\
V_{td} & V_{ts} & V_{tb} & V_{tb'}\\
V_{t'd}& V_{t's}& V_{t'b} &V_{t'b'}\\
\end{array} \right )
\end{equation}
where $V_{qb'}$ and $V_{t'q}$ are the $4\times 4$ mixing matrix elements of
the fourth generation SM and rest elements are the usual CKM matrix. 
In this mote, we reviwe our works on the SM4 and put the constraints of the
fourth generation mixing matrix elements from rare meson and lepton decays.

\section{Constriants of some 4th generation quark CKM elements}

\subsection{Constriants of $V^{*}_{t's} V_{t'b}$ from $B\to X_s
\gamma$\cite{huo1}}
  The rare decay $B\rightarrow X_s \gamma$ plays an 
 important role in present day phenomenology. The effective Hamiltonian for $B\to X_s\gamma$ 
at scales   $\mu_b={\cal O}(m_b)$ is 
    \begin{equation} \label{Heff_at_mu}
        {\cal H}_{\rm eff}(b\to s\gamma) = 
           - \frac{G_{\rm F}}{\sqrt{2}} V_{ts}^* V_{tb}
           \left[ \sum_{i=1}^6 C_i(\mu_b) Q_i + 
           C_{7\gamma}(\mu_b) Q_{7\gamma}
          +C_{8G}(\mu_b) Q_{8G} \right]\,,
    \end{equation} 
  where  the   magnetic--penguin operators
    \begin{equation}\label{O6B}
      Q_{7\gamma}  =  \frac{e}{8\pi^2} m_b \bar{s}_\alpha \sigma^{\mu\nu}
          (1+\gamma_5) b_\alpha F_{\mu\nu},\qquad           
       Q_{8G}     =  \frac{g}{8\pi^2} m_b \bar{s}_\alpha \sigma^{\mu\nu}
           (1+\gamma_5)T^a_{\alpha\beta} b_\beta G^a_{\mu\nu}             
    \end{equation}
  
 The leading logarithmic calculations can be summarized in a
  compact form  as follows \cite{Buras}:
    \begin{equation}\label{main}
      R_{{\rm quark}} =\frac{Br(B \to X_s \gamma)}
       {Br(B \to X_c e \bar{\nu}_e)}=
     \frac{|V_{ts}^* V_{tb}^{}|^2}{|V_{cb}|^2} 
     \frac{6 \alpha}{\pi f(z)} |C^{\rm eff}_{7}(\mu_b)|^2\,,
    \end{equation}
  where 
    \begin{equation}\label{g}
      f(z) = 1 - 8z + 8z^3 - z^4 - 12z^2 \ln z           
       \quad\mbox{with}\quad
        z =
       \frac{m^2_{c,pole}}{m^2_{b,pole}}
       \end{equation}
  is the phase space factor in $Br(B \to X_c e \bar{\nu}_e)$ and
  $\alpha=e^2/4\pi$. 
   In the case of four generation there is an additional contribution to $B\rightarrow X_s\gamma$ 
from the virtual exchange of the 
fourth generation up quark $t^{'}$. The Wilson coefficients of the dipole operators are given by
   \begin{equation}
      C^{\rm eff}_{7,8}(\mu_b)=C^{\rm (SM)\rm eff}_{7,8}(\mu_b)
      +\frac{V^{*}_{t^{'}s}V_{t^{'}b}}{V^{*}_{ts}V_{tb}}C^{(4)
      {\rm eff}}_{7,8}(\mu_b),
    \end{equation}
 where $C^{(4){\rm eff}}_{7,8}(\mu_b)$ present the contributions of $t^{'}$ to the Wilson coefficients, and 
$V^{*}_{t^{'}s}$ and $V_{t^{'}b}$ are two elements of the $4\times 4$ CKM matrix which now contains nine
paremeters, i.e., six angles and three phases. We recall here that the CKM coefficient 
corresponding to the $t$ quark contribution, i.e., $V_{ts}^*V_{tb}$, is factorized in 
the effective Hamiltonian. The formulas for 
calculating the Wilson coefficients $C_{7,8}^{(4)}(m_W)$ are same as their counterpaters in 
 the SM except exchanging  $t^{'}$ quark not $t$ quark and the corresponding 
 Fenymann figuers are shown in fig. 1.

  With these Wilson coefficients and the experiment results of the decays of
 $B\rightarrow X_{s}\gamma$ and $Br(B \to X_c e \bar{\nu}_e)$ \cite{data}, we obtain 
 the results of the fourth generation CKM factor $V^{*}_{t^{'}s}V_{t^{'}b}$. 
 There exist two cases, a positive factor and a negative one: 
    \begin{eqnarray}
      V^{*}_{t^{'}s}V_{t^{'}b}^{(+)} &=& [C^{(0){\rm eff}}_{7}(\mu_b)
        - C^{\rm (SM)\rm eff}_{7}(\mu_b)]
       \frac{V_{ts}^* V_{tb}}{C^{(4){\rm eff}}_{7}(\mu_b)} \nonumber \\
        &=&
      [\sqrt{\frac{R_{\rm quark}|V_{cb}|^2\pi f(z)}{
        |V_{ts}^* V_{tb}|^2 6 \alpha}}-C^{\rm (SM)\rm eff}_{7}(\mu_b)]
        \frac{V_{ts}^* V_{tb}}{C^{(4){\rm eff}}_{7}(\mu_b)}
    \end{eqnarray}
    \begin{equation}
    V^{*}_{t^{'}s}V_{t^{'}b}^{(-)}=[-\sqrt{\frac{R_{\rm quark}|V_{cb}|^2\pi 
    f(z)}{|V_{ts}^* V_{tb}|^2 6 \alpha}}-C^{\rm (SM)\rm eff}_{7}(\mu_b)]
    \frac{V_{ts}^* V_{tb}}{C^{(4){\rm eff}}_{7}(\mu_b)}
   \end{equation}
as in tab. 2,

\begin{table}[htb]
\begin{center}
\begin{tabular}{|c|c|c|c|c|c|c|c|c|}
\hline
 $m_{t^{'}}$(Gev) & 50 & 100 & 150 & 200 &250 &300 &400 \\
\hline
\hline
$V^{*}_{t^{'}s}V_{t^{'}b}^{(+)}\times 10^{-2}$&$-11.591$&$-9.259$&
$-8.126$&$-7.501$&$-7.116$&$-6.861$&$-6.548$ \\
\hline
$V^{*}_{t^{'}s}V_{t^{'}b}^{(-)}\times 10^{-3}$&$3.5684$&$2.8503$&$2.5016
$&$2.3092$&$2.191$&$2.113$&$2.016$ \\
\hline
\end{tabular}
\end{center}
\caption{The values of $V^{*}_{t^{'}s}\cdot V_{t^{'}b}$ due to masses of 
$t^{'}$ for $Br(B\rightarrow X_{s}\gamma)=2.66\times 10^{-4}$}
\end{table}
In the numerical calculations we set $\mu_b=m_b=5.0GeV$  and take the 
$t^{'}$ mass value of 50GeV, 100GeV, 150GeV, 200GeV, 250GeV, 300GeV, 400 Gev.

The CKM matrix elements obey unitarity constraints, which states that any pair of rows, or
any pair of columns, of the CKM matrix are orthogonal.  This leads to six orthogonality
conditions \cite{Ali}.  The one relevant to $b\rightarrow s\gamma$ is 
\begin{equation}
\sum\limits_{i}V_{is}^{*}V_{ib}=0, 
\end{equation} 
i.e.,
\begin{equation} V_{us}^{*}V_{ub}+V_{cs}^{*}V_{cb}+V_{ts}^{*}V_{tb}
+V_{t^{'}s}^{*}V_{t^{'}b}=0. \end{equation} 
We take the average values of the SM CKM matrix
elements  from Ref. \cite{data}.  The sum of the first three terms in eq. 
(10) is about $7.6\times 10^{-2}$.  If we take the value of
$V^{*}_{t^{'}s}V_{t^{'}b}^{(+)}$ given in Table 2, the result of the left of
(10) is much better and much more close to $0$ than that in SM,   because the
value of $V^{*}_{t^{'}s}V_{t^{'}b}^{(+)}$ is very close to the sum but has the
opposite sign. If we take $V^{*}_{t^{'}s}V_{t^{'}b}^{(-)}$, the result would
change little because the values of $V^{*}_{t^{'}s}V_{t^{'}b}^{(-)}$ are about
$10^{-3}$ order, ten times smaller than the sum of the first three ones in the
left of (10). Considering that the data of CKM matrix is not very accurate, we
can get the error range of the sum of these first three terms.  It is about
$\pm 0.6\times 10^{-2}$, much larger than $V^{*}_{t^{'}s}V_{t^{'}b}^{(-)}$.
Thus, the values of  $V^{*}_{t^{'}s}V_{t^{'}b}$ in the both cases satisfy the
CKM matrix unitarity constraints.

\subsection{Constraints on CKM Factor $V^{*}_{t^{'}s}V_{t^{'}d}$ in SM4
\cite{huo2}}

The following three
rare $K$ meson decays: two semi-leptonic decays $K^{+}\rightarrow \pi^{+}\nu\bar\nu$
and $K_{L}\rightarrow \pi^{0}\nu\bar\nu$, and one leptonic decay
$K_{L}\rightarrow \mu^{+}\mu^{-}$\cite{kk}  can provide certain constraints
on the fourth generation CKM factors,  $V^{*}_{t^{'}s}V_{t^{'}d}$ ,
$\mbox{Im}V^{*}_{t^{'}s}V_{t^{'}d}$ and  $\mbox{Re}V^{*}_{t^{'}s}V_{t^{'}d}$
respectively. 
 \begin{table}[htb]
\begin{center}
\begin{tabular}{|c||c|c|c|c|}
\hline
    & $Br(K^+\to\pi^+\nu\bar{\nu})$ &  $Br(K_L\to\pi^0\nu\bar{\nu})$ &$Br(K_L\rightarrow\mu^+\mu^-$)  \\
\hline
\hline
Experiment & $<2.4 \times 10^{-9} $\cite{BNL} & $<1.6\times 10^{-6}$\cite{Ada}
 &
 $(6.9\pm 0.4)\times 10^{-9}$\cite{BNL1}  \\
&$(4.2+9.7-3.5)\times10^{-10}$\cite{adl} & $<6.1\times 10^{-9}$\cite{ygr}&$(7.9\pm 0.7)\times 10^{-9}$\cite{KEK}\\
\hline
SM & $(8.2\pm 3.2)\times 10^{-11}$\cite{buc}& $(3.1\pm 1.3)\times 10^{-11}$ \cite{buc} &
 $(1.3\pm0.6)\times 10^{-9}$\cite{gab}  \\
\hline
\end{tabular}
\end{center}
\caption{ Comparison of $B(K^+\to\pi^+\nu\bar{\nu})$, $B(K_L\to\pi^0\nu\bar{\nu})$
and $B(K_L\to\pi^0\nu\bar{\nu})$ among  
the experimental values and SM predictions with maximum mixing.}
\end{table}

In the SM4, the branching ratios of the three decay modes mentioned 
above receive additional contributions from the up-type quark $t^{\prime}$ \cite{new1}
 \begin{equation}
Br(K^+\to\pi^+\nu\bar{\nu})=\kappa_+\left| \frac{V_{cd}V_{cs}^*}{\lambda}
P_0+\frac{V_{td}V_{ts}^*}{\lambda^5}\eta_tX_0(x_t)+\frac{V_{t'd}
V_{t's}^*}{\lambda^5}\eta_{t'}X_0(x_{t'})\right|^2, 
\end{equation}
 \begin{equation}
Br(K_L\to\pi^0\nu\bar{\nu})=\kappa_L\left| \frac{{\rm Im}V_{td}V_{ts}^*}
{\lambda^5}\eta_tX_0(x_t)+\frac{{\rm Im}V_{t'd}
V_{t's}^*}{\lambda^5}\eta_{t'}X_0(x_{t'})\right|^2, 
\end{equation} 
\begin{equation}
Br(K_L\to\mu\bar{\mu})_{\rm SD}=\kappa_{\mu} \left[ \frac{{\rm Re}
\left( V_{cd}V_{cs}^*\right) }{\lambda}P'_0+\frac{{\rm Re}\left( V_{td} 
V_{ts}^*\right) }{\lambda^5}Y_0(x_t)+\frac{{\rm Re}\left( V_{t'd}
V_{t's}^*\right) }{\lambda^5}Y_0(x_{t'})\right]^2.
\end{equation}
where $\kappa_+,\kappa_L,\kappa_{\mu}$,$X_0(x_t)$ , $X_0(x_{t'})$,
$Y_0(x_t)$ ,$Y_0(x_{t'})$,$P_0, P'_0$ may be found in Refs\cite{buras,Ham}. 
The QCD correction factors are taken to be $\eta_t =$ 0.985 and $\eta_{t'}=$ 1.0 \cite{new1}.

To solve the constrains of the 4th generation CKM matrix factors $V^{*}_{t^{'}s}V_{t^{'}d}$,
${\rm Im} V^{*}_{t^{'}s}V_{t^{'}d}$ and ${\rm Re} V^{*}_{t^{'}s}V_{t^{'}d}$, we must conculate
the Wilson coefficients  $X_0(x_{t'})$ and  $Y_0(x_{t'})$. They are the founctions of the mass
of the 4th generation top quark, $m_{t'}$. Here we give their numerical results according to several
values of $m_{t'}$, (see table 4)
 \begin{table}[htb]
\begin{center}
\begin{tabular}{ |c|| c| c| c |c |c| c |c |c |c| }
\hline
  $m_{t'}$(GeV)& 50  & 100 &150   &200  &250 &300   & 400  & 500 & 600
 \\ \hline \hline
 $X_0(x_{t'})$  &0.404  &0.873  &1.357 & 1.884& 2.474& 3.137 &4.703  & 6.615 & 8.887   \\
 \hline
\hline
  $Y_0(x_{t'})$  & 0.144  & 0.443 &0.833 & 1.303 &1.856&2.499   &4.027  & 5.919 & 8.179   \\
 \hline
\end{tabular}
\end{center}
\caption{Wilson coefficients $X_0(x_{t'})$, $Y_0(x_{t'})$ to $m_{t'}$   }
\end{table}
We found that the Wilson coefficients  $X_0(x_{t'})$ and  $Y_0(x_{t'})$
increase with the  $m_{t'}$. To get the largest constrain of the factors in eq.
(11), (12) and (13), we must use the little value of $m_{t'}$. Considering that
 the 4th generation particles must have the mass larger than $M_Z /2$ \cite{Mark},
we take $m_{t'}$ with 50 GeV to get our constrains of those three factors.

Then, from (11), (12) and (13), we arrive at the following constraints
\begin{equation}
 |V^{*}_{t^{'}s}V_{t^{'}d}| \leq 2\times10^{-4},
 \end{equation}
 \begin{equation}
|\mbox{Im}V^{*}_{t^{'}s}V_{t^{'}d}| \leq 1.2\times 10^{-4},
\end{equation}
\begin{equation} 
| \mbox{Re}V^{*}_{t^{'}s}V_{t^{'}d}| \leq 1.0\times 10^{-4}.
\end{equation}
For the numerical calculations, we will take $|\mbox{Im}V^{*}_{t^{'}s}V_{t^{'}d}| \leq 1.2\times 10^{-4}$.

 It is easy to check that the equation (14) obeys the
 CKM matrix unitarity constraint, which states that any pair of rows, or
any pair of columns, of the CKM matrix are orthogonal.\cite{data}.  The relevant one to those decay 
channels is 
\begin{equation} V_{us}^{*}V_{ud}+V_{cs}^{*}V_{cd}+V_{ts}^{*}V_{td}
+V_{t^{'}s}^{*}V_{t^{'}d}=0.
 \end{equation} 
Here we have taken the average values of the SM CKM matrix
elements  from Ref. \cite{data}.
Considering the fact that the data of CKM matrix is not yet very accurate, there still exists 
a sizable error for the sum of the first three terms. Using the value of
$V^{*}_{t^{'}s}V_{t^{'}d}$ obtained from eq. (14), the sum of the four terms
in  the left hand of (17) can still be close to $0$,  
 because the values of $V^{*}_{t^{'}s}V_{t^{'}d}$ are about $10^{-4}$
order, ten times smaller than the sum of the first three ones in the left of
(17).  Thus, the values of  $V^{*}_{t^{'}s}V_{t^{'}d}$ remain
 satisfying the CKM matrix unitarity constraints in SM4 within the present uncertainties.

\subsection{$V^{*}_{t^{'}b}V_{t^{'}d}$
from experimental measurements of $\Delta M_{B_d}$\cite{huo3}}

 $B^0_{d,s}-\bar B^0_{d,s}$ mixing proceeds to an excellent approximation only 
 through box diagrams with internal top quark exchanges in SM. In SM, the effective 
 Hamiltonian ${\cal H}_{\rm eff}(\Delta B=2)$ for $B^0_{d,s}-\bar B^0_{d,s}$
 mixing, relevant for scales $\mu_b={\cal O}(m_b)$ is given by\cite{Buras}
 \begin{eqnarray}
 {\cal H}_{\rm eff}^{\Delta B=2}=\frac{G_{\rm F}^2}{16\pi^2}{M_W^2}
 (V^{*}_{tb}V_{tq})^2 S_0(x_t) Q(\Delta B=2)+h.c. 
 \end{eqnarray}
where $ Q(\Delta B=2)=(\bar b_\alpha q_\alpha)_{V-A}
  (\bar b_\beta q_\beta)_{V-A }$, with $q=d,s$ for $B^0_{d,s}-\bar B^0_{d,s}$
respectively and $S_0(x_t)$ is the Wilson coefficient which is taken the form
\begin{eqnarray}
S_0 (x_t)=\frac{4x_t -11x_t^2 +x_t^3}{4(1-x_t )^2}
 -\frac{3}{2}\cdot \frac{x_t^3}{(1-x_t )^3}\cdot\ln{x},
\end{eqnarray}
where $x_t =m_t^2 /M_W^2$.
 The mass differences $\Delta M_{d,s}$ can be expressed 
 in terms of the off-diagonal element in the neutral $B$-meson mass matrix 
 \begin{eqnarray}
 \Delta M_{d,s}&=&2|M^{d,s}_{12}|\\
2m_{B_{d,s}}|M^{d,s}_{12}|  &= &|\langle \bar B_{d,s}^0|{\cal H}_{\rm eff}(\Delta B=2)
               |B_{d,s}^0\rangle . \nonumber
 \end{eqnarray}
  If we add a fourth sequential fourth generation up-like quark
$t^{\prime}$, the above equations would have some modification.
There exist other box diagrams contributed by $t^{'}$ (see fig. 2), similar to
the leading box diagrams in MSSM\cite{mssm}. 
 The mass differences $\Delta M_{d}$ in SM4 can be expressed 
 \begin{eqnarray}
 \Delta M_{d}&=&\frac{G_{\rm F}^2}{6\pi^2}{M_W^2}{m_{B_{d}}}
 (\hat B_{B_{d}}\hat F^2_{B_{d}})
 [\eta_t (V^{*}_{tb}V_{td})^2 S_0(x_t)+ \nonumber\\
 & +&\eta_{t^{'}} (V^{*}_{t^{'}b}V_{t^{'}d})^2
  S_0(x_{t^{'}})+\eta_{tt^{'}} (V^{*}_{t^{'}b}V_{t^{'}d})\cdot
  (V^{*}_{tb}V_{td}) S_0(x_t,x_{t^{'}})]
 \end{eqnarray}
  The new Wilson  coefficients $S_0(x_{t^{'}})$ 
present the contribution of $t^{\prime}$,
which  like  $S_0(x_{t})$  in eq. (19)
except exchanging $t^{\prime}$ quark not $t$ quark.
 $S_0(x_t,x_{t^{'}})$ present the contribution of a mixed $t-t^{\prime}$,
  which is taken the form\cite{sxy}
\begin{eqnarray}
S_0(x,y)&=&x\cdot y [-\frac{1}{y-x}(\frac{1}{4}+\frac{3}{2}\cdot \frac{1}{1-x}
 -\frac{3}{4}\cdot \frac{1}{(1-x)^2}\ln{x}+ \nonumber \\
 &+&(y\leftrightarrow x)-\frac{3}{4}\cdot \frac{1}{(1-x)(1-y)}]
 \end{eqnarray}
 where $x= x_t =m_t^2 /M_W^2$, $y= x_{t^{'}} =m_{t^{'}}^2 /M_W^2$. The 
 numerical results of $S_0 (x_{t^{\prime}})$ and $S_0 ({x_t}, {x_{t^{\prime}}})$
 is shown on the tab. 5.
\begin{table}[htb]
\begin{center}
\begin{tabular}{|c||c|c|c|c|c|c|c|c|c|c|}
\hline
$m_t^{'}$(GeV) &50  &100  & 150 & 200 & 250 &300 &350 &400&450&500   \\
\hline
$S_0 (x_{t^{'}})$ &0.33&1.07&2.03&3.16&4.44&5.87&7.47&9.23&11.15&13.25 \\
$S_0 (x_t ,x_{t^{'}})$ & 0.48&-7.03&-4.94&-5.09& -5.39&-5.87&-5.99&-6.25
&-6.49& -6.72 \\
\hline
\hline
$m_t^{'}$(GeV) &550  &600  & 650 & 700 &750 &800 &850 &900&950&1000   \\
\hline
$S_0 (x_{t^{'}})$ &15.52&17.97&20.60&23.41&26.40&29.57&
32.93 & 36.47 & 40.96 & 44.11  \\
$S_0 (x_t ,x_{t^{'}})$ &-6.92& -7.11&-7.28&-7.44&-7.60&-7.74& -7.87
& -7.99& -8.12 & -8.23 \\
\hline
\end{tabular}
\end{center}
\caption[]{The Wilson coefficients $S_0 (x_{t^{\prime}})$ 
and $S_0 ({x_t}, {x_{t^{\prime}}})$ to  $m_{t^{'}}$}
\end{table}

The short-distance QCD correction  factors 
$\eta_{t^{'}}$ and $\eta_{tt^{'}}$ can be
calculated like $\eta_c$ and $\eta_{ct}$ in the mixing of $K^0 - \bar K^0$,
 which the NLO values are given in refs\cite{Buras,eta}, 
relevant for scale not ${\cal O}(\mu_c)$ but ${\cal O}(\mu_b)$. 
In leading-order, $\eta_{t}$ is calculated by
\begin{eqnarray}
\eta^0_t =[ \alpha_s (\mu_t )]^{(6/23)},\ \ \ 
  \alpha_s (\mu_t) =\alpha_s (M_Z) [1+\sum^{\infty}_{n=1}
 (\beta_0 \frac{\alpha_s (M_Z)}{2\pi} {\rm In}\frac{M_Z}{\mu_t})^n],
\end{eqnarray}
with its numerical value in tab. 6. The formulae of factor $\eta_{t^{'}}$ is
similar to the above equation except for exchanging $t$ by $t^{'}$.
For simplicity, we take $\eta_{tt^{'}} =\eta_{t^{'}}$. We give the 
numerical results in tab.7.
\begin{table}[htb]
\begin{center}
\begin{tabular}{|c||c|c|c|c|c|c|c|c|c|c|}
\hline
$m_t^{'}$(GeV) &50  &100  & 150 & 200 & 250 &300 &350 &400&450&500   \\
\hline
$\eta_{t^{'}}$ &0.968&0.556&0.499&0.472&0.455&0.443&0.433&0.426&0.420&
0.416 \\
\hline
\hline
$m_t^{'}$(GeV) &550  &600  & 650 & 700 &750 &800 &850 &900&950&1000   \\
\hline
$\eta_{t^{'}}$ &0.412&0.408&0.405&0.401&0.399&0.396&
0.395 & 0.393 &0.391 &0.389  \\
\hline
\end{tabular}
\end{center}
\caption[]{The short-distance QCD factors  $\eta_{t^{'}}$,
$\eta_{t{t^{'}}}(=\eta_{t^{'}})$ to  $m_{t^{'}}$}
\end{table}
In the last of this section, we give other input parameters necessary
in this note. (See the following tab.).
\begin{table}[htb]
\begin{center}
\begin{tabular}{ |c| c| c| c | }
\hline
 $\overline m_c(m_c(pole))$ & $1.25\pm0.05$GeV & $M_W$&$80.2$GeV\\
 $\overline m_t(m_t(pole))$ & $175$GeV &$\hat F_{B_{d}}\sqrt{\hat B_{B_{d}}})$
 & $215\pm40$MeV\\
 $\Delta M_{B_d}$& $(0.473\pm0.016)(ps)^{-1}$ &$\xi_s$ & $1.14\pm0.06$  \\
 $\Delta M_{B_s}$& $>14.3(ps)^{-1}$ & $G_{\rm F}$&$1.166\times10^{-5}$GeV${^{-2}}$   \\
 \hline
\end{tabular}
\end{center}
\caption{Neumerical values of the input parameters\cite{ali}.}
\end{table}

Now, we can put  the constraints of the fourth generation CKM factor
$V^{*}_{t^{'}b}V_{t^{'}d}$ from the present experimental value
of $\Delta M_{B_d}$. 
We change the form of eq. (21) as a quadratic equation about
$V^{*}_{t^{'}b}V_{t^{'}d}$. By solving it , we can get two analytical solution
$ V^{*}_{t^{'}d}V_{t^{'}b}^{\rm (1)}$ (absolute value is the large one) and 
 $V^{*}_{t^{'}d}V_{t^{'}b}^{\rm (2)}$ (absolute value is the small one).
   However, experimentally, it is not accurate for the measurement of CKM matrix 
  element $V_{td}$\cite{Buras,data}. So, we have to search other 
 ways to solve this difficulty. Fortunately,  the  CKM unitarity 
 triangle\cite{Ali}, i.e. the graphic representation of the unitarity relation for 
 $d,b$ quarks,
 which come from the orthogonality condition on the first and
 third row of $V_{\rm CKM}$,
 \begin{eqnarray}
 V_{ud} V^{*}_{ub} + V_{cd} V^{*}_{cb} + V_{td} V^{*}_{tb} =0, 
 \end{eqnarray}
  can be conveniently depicted as a triangle relation in the complex
 plane, as shown in the following figure.
  From the above equation, we can  give the constraints of 
 $V_{td} V^{*}_{tb}$\cite{9905397},
 \begin{equation}
 0.005 \leq |V_{td} V^{*}_{tb} | \leq 0.013
 \end{equation}
Then, we give the final results as shown in the  figs. 3.

We must announce that figs. 3 only show the curves with 
$V^{*}_{t^{'}d}V_{t^{'}b}^{\rm (2)}$ (absolute value is the small one) firstly.
 Because
the absolute value of $V^{*}_{t^{'}d}V_{t^{'}b}^{\rm (1)}$  is generally larger
than 1. This is contradict to the unitarity of CKM matrix. So, we don't
think about this solution. From the figs. 3, we found all curves are in the
range from $-1\times 10^{-4}$ to $0.5 \times 10^{-4}$ when we considering 
the constraint of $V_{td} V^{*}_{tb}$. That is to say, the absolute value
of $V^{*}_{t^{'}d}V_{t^{'}b}$ is about $\sim 10^{-4}$ order. This is a
very interesting result. 

These CKM matrix elements obey unitarity constraints. 
With the fourth generation quark $t^{'}$, eq. (9) change to ,
\begin{equation}
 V_{ud}^{*}V_{ub}+V_{cd}^{*}V_{cb}+V_{td}^{*}V_{tb}
+V_{t^{'}d}^{*}V_{t^{'}b}=0. 
\end{equation} 
We take the average values of the SM CKM matrix elements from Ref. \cite{data}.
The sum of the first three terms in eq. (24) is about $\sim 10^{-2}$ order.
If we take the value of $V^{*}_{t^{'}s}V_{t^{'}b}^{\rm (2)}$ the result of the 
left of (26) is better and more close to $0$ than that in SM, when
 $V^{*}_{t^{'}s}V_{t^{'}b}^{(2)}$ takes negative values. Even if 
 $V^{*}_{t^{'}s}V_{t^{'}b}^{(2)}$ takes positive values, the sum of (26) 
 would change very little because the values of 
 $V^{*}_{t^{'}d}V_{t^{'}b}^{\rm (2)}$ are about $10^{-4}$ order, two orders 
 smaller than the sum of the first three ones in the left of (24).
Considering that the data of CKM matrix is not very accurate, we can 
get the error range of the sum of these first three terms.  It is  much larger
than $V^{*}_{t^{'}d}V_{t^{'}b}^{\rm (2)}$. Thus,in the case 
 the values of  $V^{*}_{t^{'}d}V_{t^{'}b}$ satisfy the CKM matrix unitarity 
 constraints.

\vskip 1.5cm

We can see the  order of these 4th generation CKM
matrix elements, such as $V^{*}_{t^{'}d}V_{t^{'}b}$ doesn't contradict to the
hierarchy of the CKM matrix elements or the quarks mixing 
angles\cite{Wakaizumi,hierarchy}. Moreover,  it seem to prove the hierarchy.
The hierarchy in the quarks mixing angles is  clearly presented in the
Wolfenstein parameterization\cite{wolfenstein} of the  CKM matrix. Let's see
CKM matrix firstly, 
\begin{equation}
V_{\rm CKM} = \left (
\begin{array}{lcrr}
V_{ud} & V_{us} & V_{ub} & \cdots\\
V_{cd} & V_{cs} & V_{cb} & \cdots\\
V_{td} & V_{ts} & V_{tb} & \cdots\\
\vdots& \vdots& \vdots &\ddots\\
\end{array} \right )
\sim \left (
\begin{array}{lcrr}
1 & \lambda & \lambda^3 & \cdots\\
-\lambda& 1 & \lambda^2 & \cdots\\
\lambda^3 & -\lambda^2 & 1&\cdots\\
\vdots& \vdots& \vdots &\ddots\\
\end{array} \right )
\end{equation}  
with $\lambda=\sin^2 \theta=0.23$. Now, the hierarchy can be expressed in
powers of $\lambda$. We found, the magnitudes of the mixing angles are about 1 among
the $same$ generations, $V_{ud}$, $V_{cs}$ and $V_{tb}$. For different generations,
the magnitudes are about $\lambda$ order between $1st$ and $2nd$ generation,
$V_{us}$ and $V_{cd}$, as well as about $\lambda^2$ order 
between $2nd$ and $3rd$ generation, $V_{cb}$ and $V_{ts}$. The magnitudes 
are about $\lambda^3$ order between the $1st$ and $third$ generation,
$V_{ub}$ and $V_{td}$. Then, there should be an interesting problem: If the fourth
generation quarks exist, how to choose the  order do the magnitude of the mixing
 angles concern the fourth generation quarks? Because there is not direct experimental
 measurement of the fourth generation quark mixing angles, one have to look for other
 indirect methods to solve the problem.
Many refs. have already talked about these additional CKM mixing 
angles\cite{vec-like,sterile,McKay,exp}, like the 
vector-like quark models\cite{vec-like}, the four neutrinos 
models\cite{sterile} and 
the sequential four generations models\cite{McKay}. For simple,
 we give a guess for the magnitude of the fourth generation mixing angles. Similar to
 the general CKM matrix elements magnitude order, the fourth generation
 ones are about $\lambda^4 \sim \lambda^5$ order between the $1st$ and $4th$ 
 generation, such as $V_{t^{'} d}$, as well as $\lambda^2 \sim \lambda^3$ between
 the $2nd$ and $4th$ generation, such as $V_{t^{'} s}$. For the mixing between 
 the $3rd$ and $4th$ generation quarks, such as $V_{t^{'} b}$, we take the magnitude 
 as 1 because the mass of the fourth generation quark $t^{'}$ is the same order,
 $10^2$, as the top quark $t$. So $V_{t^{'} b}$ should take the order of $V_{tb}$.
 Then, the magnitude order of the fourth generation CKM factor 
 $V^{*}_{t^{'}d}V_{t^{'}b}$ is about $\lambda^4 \sim \lambda^5$, i.e. 
 $< \lambda^4$. From figs. 3, we found that
 the  numerical results, $V^{*}_{t^{'}d}V_{t^{'}b}^{(2)}$, satisfy this guess.
  At last, the factor $V^{*}_{t^{'}d}V_{t^{'}b}$ constrained from 
 $\Delta M_{B_d}$ does not contradict to the CKM matrix texture. Moreover, it
 seem to support the existence of the fourth generation.

\section{Conclusion}

In summary, we study the constraints of some 4th generation quark mixing
matrix from rare $K, B$ decays. We find they satisfy the unitarity conditions
of the CKM matrix. We also talk about the texture of the fourth generation CKM
matrix. All these constriants could provide a possible signal of new physics.

\section*{Acknowledgments}

This research is supported  by the the Chinese Postdoctoral 
Science Foundation and CAS K.C. Wong Postdoctoral Research Award  Fund.

\newpage
\begin{figure}
\epsfxsize=10cm
\epsfysize=8cm
\centerline{
\epsffile{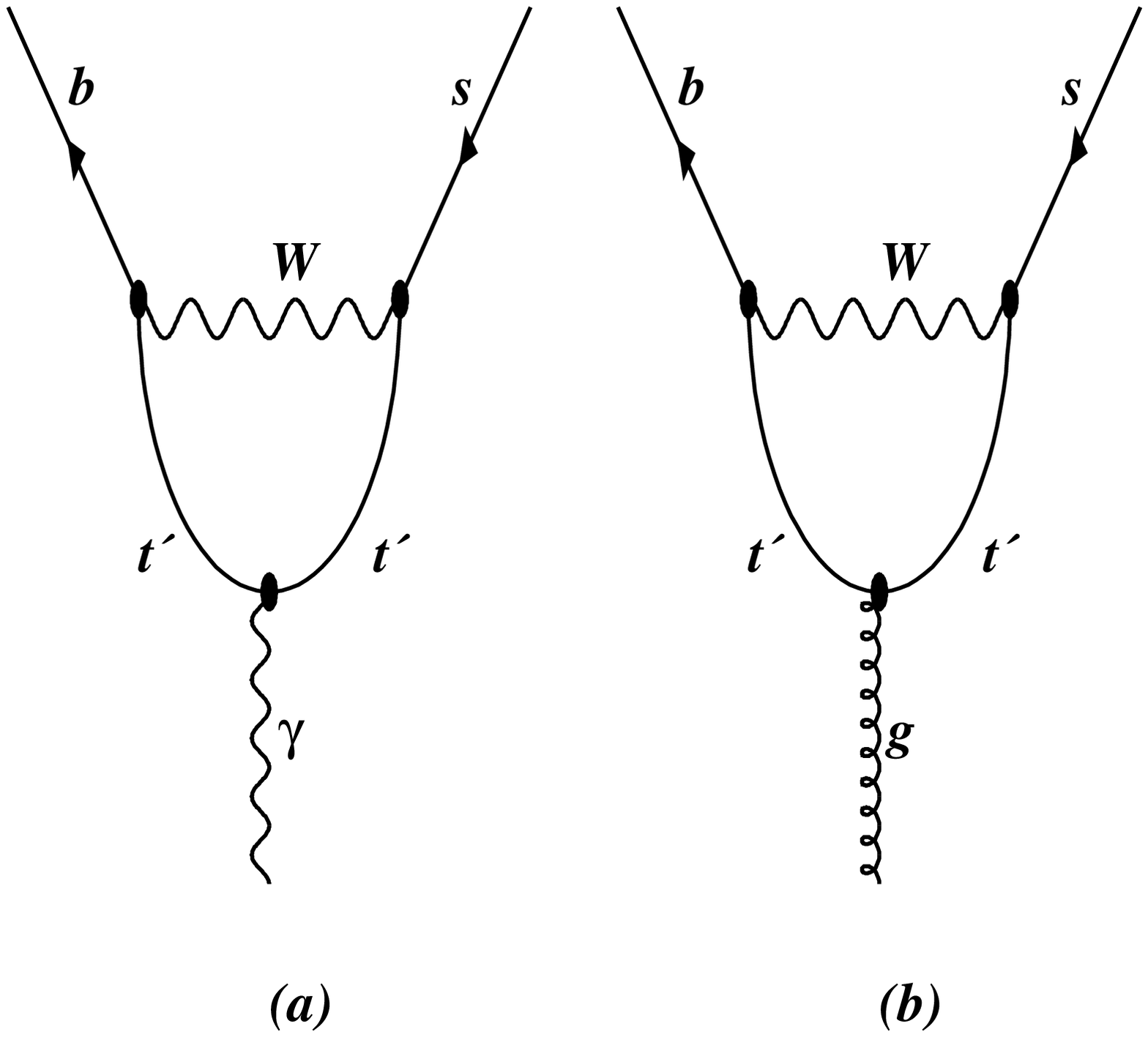}}
\vskip 0.0cm
\caption{Mabnetic Photon (a) and Gluon (b) Penguins with $t^{'}$.}
\end{figure}

\newpage
\begin{figure}
\epsfxsize=20cm
\epsfysize=18cm
\centerline{
\epsffile{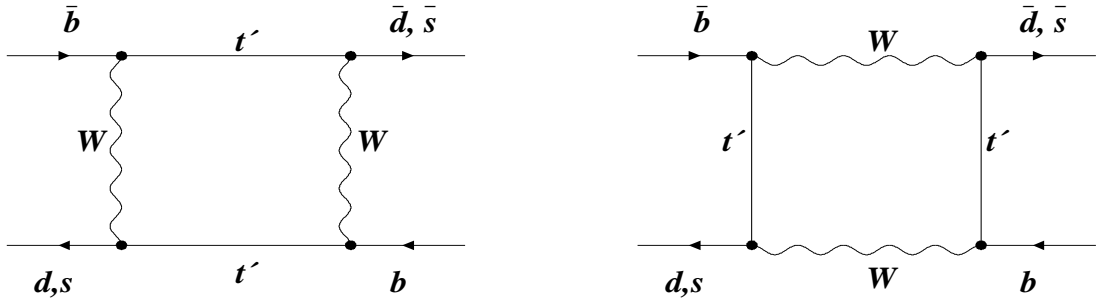}}
\vskip -10.0cm
\caption{The Additional Box Diagrams to $B^0_{d,s} -\bar B^0_{d,s}$
with the fourth up-like quark $t^{'}$.}
\end{figure}

\newpage
\begin{figure}
\epsfxsize=10cm
\epsfysize=10cm
\centerline{
\epsffile{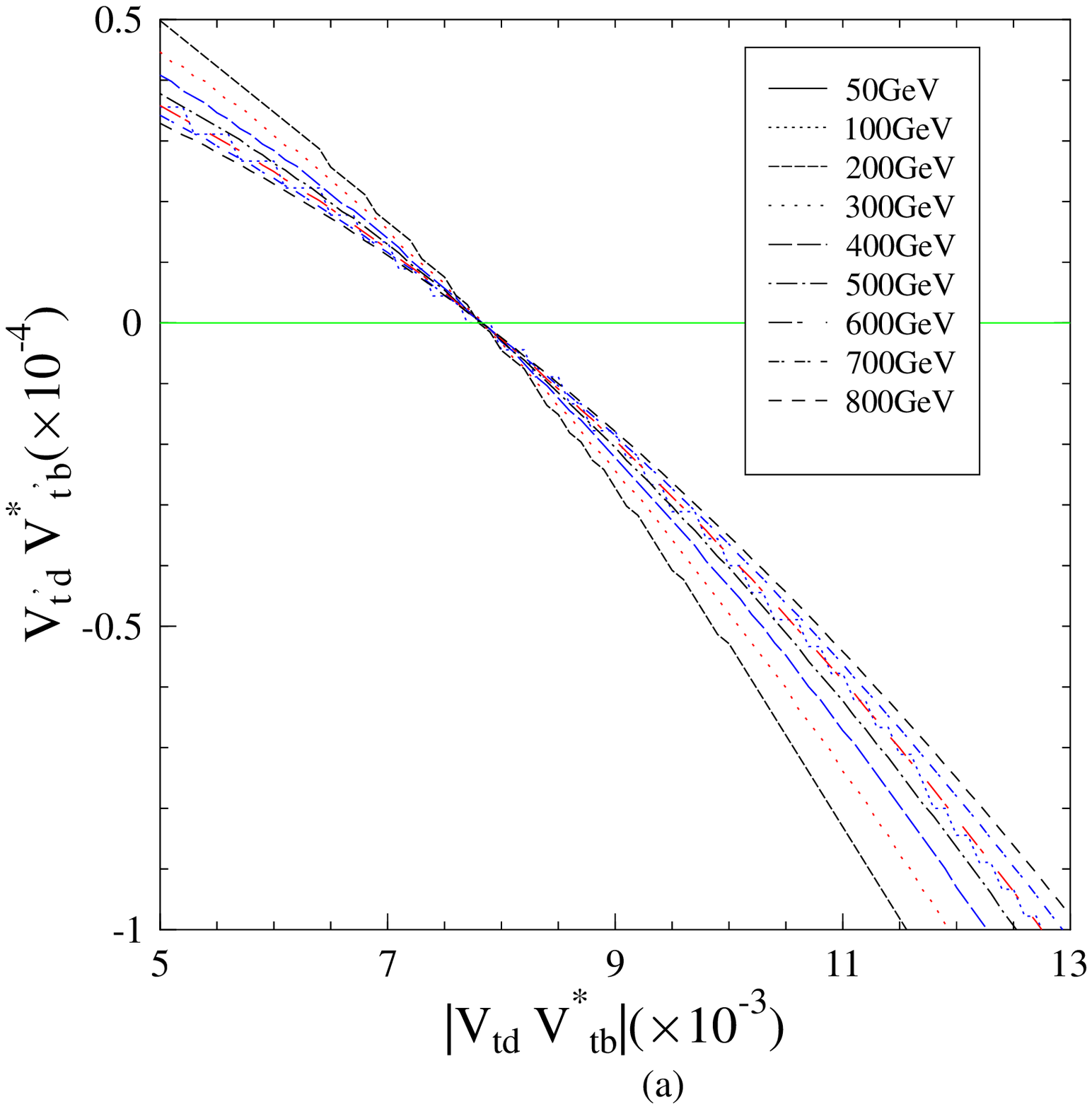}}
\vskip 0cm
\end{figure}

\begin{figure}
\epsfxsize=10cm
\epsfysize=10cm
\centerline{
\epsffile{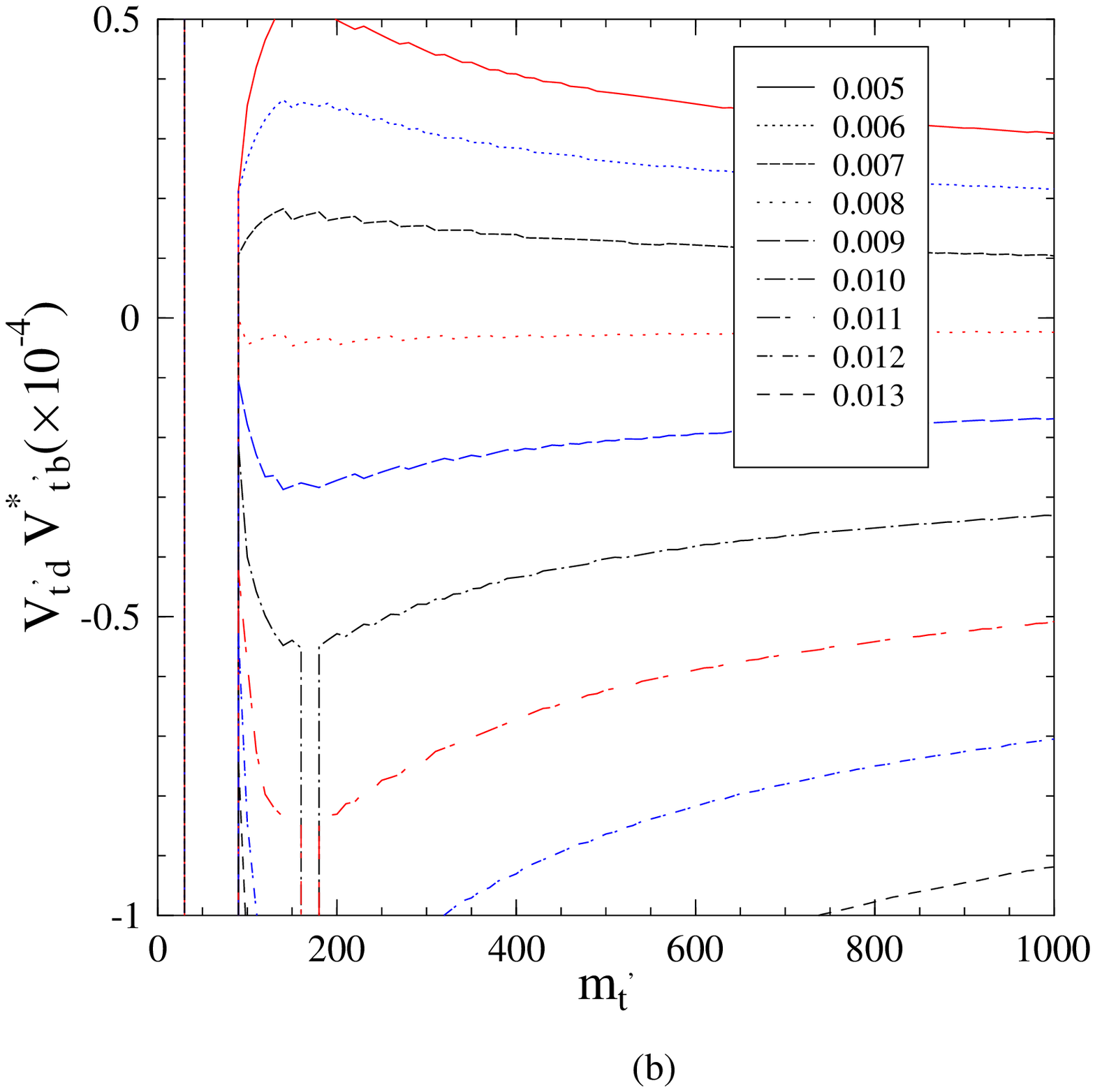}}
\vskip 0cm
\caption{Constraint of the 4th generation CKM factor
$V^{*}_{t^{'}d}V_{t^{'}b}$ to (a) $|V_{td} V^{*}_{tb} |$ with 
 $m_{t^{'}}$ range from 50GeV to 800GeV, (b) to
  $m_{t^{'}}$ with $|V_{td} V^{*}_{tb}|$
range from 0.005 to 0.013.}
\end{figure}

\begin{thebibliography}{99}
\bibitem{grand}P. Langacker, Phys. Rep. {\bf 72}, No. 4, (1981) 185.
\bibitem{supers}M.F. Sohnius, Phys. Rep. {\bf 128}, No. 2\&3 (1985) 39.

\bibitem{e821}Muon g-2 Collaboration, H.N. Brown {\it et al.}, Phys. Rev.
Lett. {\bf 86}, 2227 (2001).
\bibitem{neutrino}Y. Fukuda {\it et al.}, Phys. Lett. {\bf B436} 33 (1998);
Phys. Rev. Lett. {\bf 81}, 1562 (1998).

\bibitem{Mark} G.S.  Abrams et al., Mark II Collab., Phys.  Rev. Lett.  {\bf 63}
(1989) 2173; B.  Advera et al.,
 L3 Collab.,  Phys.  Lett.  B {\bf 231} (1989) 509;I.  Decamp et al.,
 OPAL Collab.,  {\it ibid.}, {\bf 231} (1989) 519; M.Z.  Akrawy et al.,
 DELPHI Collab., {\it ibid.}, {\bf 231} (1989) 539;
 C.Caso et al., (Particle Data Group), Eur.  Phys.  J.C {\bf 3} (1998) 1.  
\bibitem{Berez}Z.  Berezhiani and E, Nardi, Phys.  Rev.  D {\bf 52} (1995) 3087; 
 C.T.  Hill, E.A.  Paschos, Phys.  Lett.  B {\bf 241} (1990) 96.  
\bibitem{vec-like}Y. Nir and D. Silverman, Phys. Rev. {\bf D42} (1990) 1477;
  W-S, Choong and D. Silverman, Phys. Rev. {\bf D49} (1994) 2322; L.T. Handoko, Hep-ph/9708447.
\bibitem{sterile}V. Barger, Y.B. Dai, K. Whisnant and B.L. Young, Hep-ph/9901380;
R.N. Mohapatra, hep-ph/9702229; S. Mohanty, D.P. Roy and U. Sarkar, hep-ph/9810309;
S.C. Gibbons,{\it et al}., Phys. lett. {\bf B430} (1998) 296;
V. Barger, K. Whisnant and T.J. Weiler, Phys. lett. {\bf B427}, (1998) 97;
V. Barger, S. Pakvasa, T.J. Weiler and K. Whisnant, Phys. Rev. {\bf D58} (1998) 093016.
\bibitem{McKay}J.F.  Gunion, Douglas W. McKay, H.  Pois, Phys.  Lett.  B {\bf 334} (1994) 339;
 Phys.  Rev.  D {\bf 51} (1995) 201.

\bibitem{huo1}C.S. Huang, W.J. Huo and Y.L. Wu, Mod. Phys. Lett. {\bf A14},
(1999) 2453.

\bibitem{Buras}Andrzej J.  Buras; hep-ph/9806471.

\bibitem{data}C.Caso et al., (Particle Data Group), Eur. Phys. J. C{\bf 3}
(1998) 1; B.  Grinstein, M.J.  Savage, M.B.  Wise, Nucl.  Phys.
 B{\bf 319} (1998) 271.

\bibitem{Ali}A.  Ali, hep-ph/9606324; hep-ph/9612262. 

\bibitem{huo2} C.S. Huang, W.J. Huo and Y.L. Wu, Phys.Rev. {\bf D64} (2001)
016009.

\bibitem{kk}R. D. Peccei, hep-ph/9909236;
 T. Hattori, T. Hasuike and S. Wakaizumi, hep-ph/9808412; A.J. Buras, hep-ph/9901409.


\bibitem{BNL}S. Adler, {\it et al}., Phys. Rev. Lett. {\bf B76} (1996) 1421.

\bibitem{Ada}J. Adams,{it et al},. hep-ex/9806007.

\bibitem{BNL1}A.P. Heinson, {\it et al}., Phys. Rev. {\bf D51} (1995) 985.

\bibitem{ygr}Y. Grossman and Y. Nir. Phys. Lett. {\bf B398} (1997) 163.

\bibitem{KEK}T. Akagi, {\it et al}., Phys. Rev. {\bf D51} (1995) 2061.

\bibitem{gab}F. Gabbiani, hep-ph/9901262.

\bibitem{buc}G. Buchalla, A.J. Buras, hep-ph/9901288.

\bibitem{adl}S. Adler, {\it et al}., Phys. Rev. Lett. {\bf B79} (1997) 2204.

\bibitem{new1}T. Hattori, T. Hasuike, S. Wakaizumi, hep-ph/9804412.

\bibitem{buras}A.J. Buras, hep-ph/9806471.

\bibitem{Ham} G. Buchalla, A.J, Buras, M.E. Lautenbacher, Rev. of Mod. Phys.
{\bf 68} (1996) 1125 and references therein; 
 E. A. Paschos, Y.L. Wu, Mod. Phys. Lett. {\bf A6} (1991) 93.

\bibitem{huo3}C.S. Huang, W. J. Huo and Y.L. Wu,  hep-ph/0006110.

\bibitem{mssm}I. Hinchliffe and N. Kersting, hep-ph/0003090.

\bibitem{sxy}J.F. Donoghue, E. Golowich and B.R. Holstein, {\it
Dynamics of the Standard Model} (Cambridge University Press, New York, 1992).


\bibitem{eta}S. Herrich and U. Nierste, hep-ph/9604330; hep-ph/9310311; 
S. Herrich, hep-ph/9609376.

\bibitem{ali}A. Ali and D. London, hep-ph/0002167.

\bibitem{9905397}G. Barenboim, G. Eyal and Y. Nir, hep-ph/9905397.
\bibitem{hierarchy}M. Leurer, Y. Nir, N. Seiberg, Nucl. Phys. 
{\bf B420} (1994) 468; P. Kielanowski, {\it et al}, hep-ph/0002062.

\bibitem{wolfenstein}L. Wolfenstein, Phys. Rev. Lett. {\bf 51} (1983) 1945.
\bibitem{Wakaizumi}T. Hattori, T. Hasuike and S. Wakaizumi, Phys. Rev. {\bf
D60} (1999) 113008.

\bibitem{exp}${\rm D\O}$ Collab., S. Abachi et al., Phys. Rev. Lett. {\bf 78} (1997) 3818.

\end{thebibliography}
\end{document}